# CDTOM: A Context-driven Task-oriented Middleware for Pervasive Homecare Environment

Hongbo Ni[1, 2], Bessam Abdulrazak[2], Daqing Zhang[3], Shu Wu[2]

[1]School of Computer Science, Northwestern Polytechnic University, China
`nihb@nwpu.edu.cn`
[2]Department of Computer, University of Sherbrooke, Canada
`{Bessam.Abdulrazak,Shu.wu}@usherbrooke.ca`
[3]Handicom Lab, Institut Telecom SudParis, France
`Daqing.Zhang@it-sudparis.eu`

*ABSTRACT*

*With the growing number of the elderly, we see a greater demand for home care, and the vision of pervasive computing is also floating into the domain of the household that aims to build a smart home which can assist inhabitants (users) to live more conveniently and harmoniously. Such health-care pervasive applications in smart home should focus on the inhabitant's goal or task in diverse situations, rather than the various complex devices and services. The core challenge for homecare design is to perceive the environment and assess occurring situations, thus allowing systems to behave intelligently according to the user's intent. Due to the dynamic and heterogeneous nature of pervasive computing environment, it is difficult for an average user to obtain right information and service and in right place at right time. This paper proposes a context-driven task-oriented middleware (CDTOM) to meet the challenge. The most important component is its task model that provides an adequate high-level description of user-oriented tasks and their related contexts. Leveraging the model multiple entities can easily exchange, share and reuse their knowledge. Based on the hierarchy of task ontology, a novel task recognition approach using CBR (case-based reasoning) is presented and the performance of task recognition is evaluated by task number, context size and time costing. Moreover, a dynamic mechanism for mapping the recognized task and services is also discussed. Finally, we present the design and implementation of our task supporting system (TSS) to aid an inhabitant's tasks in light of his lifestyle and environment conditions in pervasive homecare environment, and the results of the prototype system show that our middleware approach achieves good efficiency of context management and good accuracy of user's activity inference, and can improve efficiently quality of user's life.*

*KEYWORDS*

*Pervasive computing, Homecare, Task, Context-driven, ECA Rule*

## 1. INTRODUCTION

With the increasing number of aging population, more people will be suffering from age-related ailments such as memory loss, chronic illnesses, renal disorders, and communication difficulties with others. Around 80% of people over 60 have a visual impairment, 75% of people over 60n have a hearing impairment and 22% have both a hearing and visual impairment [1]. These disabilities can reduce the ability of older people to live independently, resulting in a greater need for personal care and assistive technology and older adults may also have difficulty understanding language spoken in noisy conditions or at fast rates [2]. Many age related medical conditions are of a longer term degenerative rather than an acute nature, where the condition increases in severity with age.

Thus care for the elders will increasingly be delivered within a domestic setting rather than in an acute care facility or hospital environment, so effective integration and sharing at the point of use of accurate and timely personal and social care will be essential.

Smart home has been always considered as an important test-bed of the pervasive computing. The home system will effectively integrate communication and computing networks among the previously separated equipments in home and incorporate core tenets of pervasive computing. If the home system intends to dynamically adapt its behaviors according to the user's activities and environments, awareness of the user's activities and environment is required. Given the need for smart home applications to be smarter in assisting their inhabitants, this paper presents our design and implementation of a context-driven task-oriented middleware (CDTOM) for Pervasive Homecare, which is established in such a fact that man is a creature of habit, and he will perform a certain activity at a particular situation as a routine. The CDTOM uses the abstraction of tasks in order to separate logical relations of relevant items from the services realizing and fulfilling the intended goals. This CDTOM is able to support human requirements and preferences better because of the following reasons:

(1) Task definition models human preferences and requirements better than service-orientation models adopted in earlier works;

(2) Separation of tasks and services would allow for greater flexibility of changing the tasks without changing the services and vice-verse;

(3) It hides the complexity of compositing various services in pervasive environment from the users.

CDTOM provides a number of system-level services, such as context data acquisition, context storage, context-driven task reasoning, service discovery and task-oriented mapping, to facilitate the development and deployment of various pervasive homecare applications. The middleware architecture separates context management, context-driven task, and task-related services in different layers that are accessible by the applications. This scheme, not only decouples the dependency of techniques used in individual layers, but also provides a greater flexibility for the selection and deployment of appropriate techniques in each layer by the specific system requirements. Similarly, the development effort and cost of Homecare system would be reduced through the task layer and its programming interfaces.

The rest of the paper is organized as follows. Section 2 gives a brief review of relevant work. Section 3 introduces the architecture of CDTOM, and also its key components. Section 4 introduces the context-driven task concept and model, followed by the details of construction of the modelling approach using OWL in Section 4. In Section 5, the inference algorithm and the evaluation results of elders' tasks with case-based reasoning (CBR) are discussed. The service mapping mechanism in CDTOM is briefly described in Section 6, and the system prototype of CDTOM is presented in Section 7. We conclude the paper in Section 8..

## 2. RELATED WORK

A number of famous projects have been done in the area of pervasive homecare environment. The Aware Home [3] in GaTech was constructed to provide several services to its residents which can enhance their quality of life. Extended from Robotic Room2, Sensing Room [4] was developed in the University of Tokyo, which can imitate a real-life within a single room. EasyLiving Project [5] at Microsoft Research aimed to develop prototype architecture and technologies for constructing intelligent environments. Mavhome (Cook, et al, 2003) in UTA is noteworthy and successful, which combined technologies from artificial intelligence, machine learning, and databases to create a smart home that acts as an agent-based software system. The INHOME project [7] aims at providing the means for improving the quality of life of elderly people at home, by developing technologies for managing their domestic environment and enhancing their autonomy and safety at home (e.g., activity monitoring, simple home environment management, flexible AV streams handling, flexible household appliance access). Among more recent works, the "ubiquitous home" is a real-life test-bed for home-based context-aware service experiments [8] in Japan. A set of implemented context-aware services has been evaluated by means of real-life experiments with elderly people. Pervasive self care is a conceptual framework for the development of pervasive self-care services [9]. This study has been promoted in the framework of self care, an initiative by the Department of Health in the UK that aims at treating patients with long-term conditions near home. The authors of Semantic Space [10] and SOCAM [11] have studied the context data management over physical spaces in different domains. Judd and Steenkiste proposed the Contextual Information Service (CIS) [12] that defines physical spaces as one type of context source while the other types are people, devices and networks. CAMPH project proposes a Context-Aware Middleware for elder-care in pervasive environment [13].

These projects try to build an intelligent environment as one that is able to acquire and apply knowledge about its inhabitants and surroundings in order to assist the users and meet the goal of comfort and efficiency. But most of them have some limitations in assisting an average user to perform his/her tasks with diverse and configurable services and resources.

Another related research subject is task computing. There are a number of related works in task computing focusing on context-aware task modeling. However, most of them have been centered on the physical aspects of the user context (e.g. number, time, location) and the environment context (device proximity, lighting condition) [13][14]. This is despite the fact that many authors have long recognized the importance of using the cognitive aspect of the user context (such as users' goals, preferences and emotional state etc.). To date, very little work has been done to develop and apply such models in building ubiquitous computing application [15] [16]. Moreover, there are also many important task-based projects in pervasive computing community, for example, the task-driven computing [17] and task computing [18]. However there are some weaknesses in them. On the one hand, task-driven computing has treated a task as merely a set of applications, and any progress through the task is modeled through the application state. The fundamental problem of it is that it is focused on the wrong item; applications are only the tools used to carry out a task; they do not necessarily represent the task itself. We choose not to look beyond the applications, but focus on a user's goals. On the other hand, task computing is a user oriented framework that lets users accomplish complex tasks on the fly from open, dynamic and distributed set of resources. All resources within task computing are realized as services available to the task computing clients. The compositions can then be executed as atomic services or can be used for further compositions, but this is a time consuming activity and usually means that setting up and becoming familiar with the environment takes longer than performing the intended tasks. This also means the user must have a reasonably in-depth knowledge of how to configure services based on his/her requirements.

International Journal of UbiComp (IJU), Vol.2, No.1, January 2011

Compared to all these previous techniques, our proposed context-driven task supporting solution in CDTOM is to highlight the importance of the contexts related to a task, and also represent users' goals in tasks and bridge the gap between tasks and available services.

## 3. OVERVIEW OF THE CDTOM ARCHITECTURE

The overall architecture of CDTOM is illustrated in Fig. 1, which is to support the context-driven tasks in pervasive homecare environment, and provide the assistance of daily activities
and necessary healthcare to the elders. The CDTOM consists of the following four logical layers:

- Context Provider: Real-world contexts often originate from diverse sources, leading to dissimilar approaches to generating context description. Context providers obtain raw context information from various sources such as hardware sensors and software programs and transform them into context mark-ups. Some context providers, including the location context providers, the environment context providers (which gather environmental information such as temperature, noise, and light from embedded sensors), work with the hardware sensors deployed in our prototypical smart space. Software-based context providers include the task context providers, which extracts schedule information from the inhabitant's PDA or WorkStation. We implemented these providers as Universal Plug and Play (www.upnp.org) services that can dynamically join a smart space, obtain IP addresses, and multicast their presence.

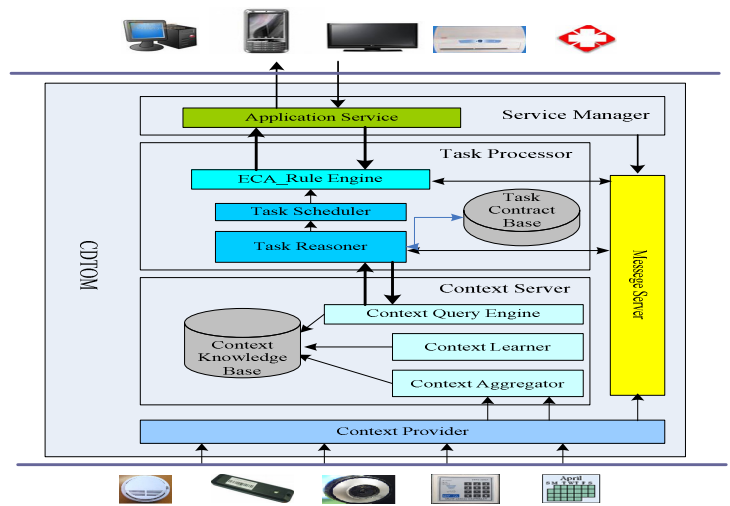

Figure 1. Architecture and Key Components of CDTOM

- Context Server: The main objective of this layer is to enable the effective and efficient context data aggregator, storage and query. Its functions are deployed in the system server. Briefly, the context aggregator component discovers context providers and gathers context mark-ups from them. The context knowledge Base (CKB) provides persistent context knowledge storage. Contexts in smart spaces display very high change rates, so the aggregator must regularly update the CKB with fresh contexts. The aggregator monitors the wrappers' availability and manages the scope of contexts in the CKB. When a context

International Journal of UbiComp (IJU), Vol.2, No.1, January 2011

wrapper joins the smart space, the context aggregator adds the provided contexts to the CKB, and when the wrapper leaves, the aggregator deletes the contexts it supplied to avoid stale information. The layer also implements a declarative, SQL-based context query interface for services or applications to acquire context data or subscribe to event notifications from physical environment. Besides, context learner is to track and update the elder's preference with implicitly learning method, and store the revision into CKB.

- Task Processor: this is the most important layer in CDTOM; there are three key components in it: Task Reasoner, Task Scheduler, and the Rule Engine. The task reasoned is to infer the inhabitant's current task according to the real-time context data and the historical context information. The Task Scheduler enables the high-priority task, e.g. fire-alarm, to be executed in a priority-scheduling algorithm when there are several concurrent tasks in a given context. The ECA rule engine is to trigger the execution the inferred task by available and competent services and resources based on the task contract and current contexts. The services will be invoked by Event-Condition-Action (ECA) mechanism.

- Service Manager: This logical layer of CDTOM manages context-aware services to enable the orchestration of homecare applications, and also it is the coordinator between the inferred tasks and diverse services enabling the task to be executed automatically and adaptively. All the services in smart home will be registered and updated as a bundle in the service manager located in the OSGi service gateway. It is implemented by a mapping scheme for abstract and specific service management. The abstract view of the scheme enables the description of a virtual service type, e.g. media rendering, while the specific view allows a physical service to be invoked adaptively according to the current contexts, e.g. CD player or Smart TV.

## 4. CONTEXT-DRIVEN TASK MODEL (CDTM)

In pervasive computing environment, a user's task is usually dependent to the current contexts, and this relationship can be captured in our context-driven task model definition. The key point of the model is on the context information that surrounds performing a task by a user. The crucial components in the model are environment, a user and his/her tasks. From our point of view, the environment includes all the resources around the user, and the state of the user and the environment constitute a context surrounding a task. Here, the task can be represented as a description of a goal to be achieved by the user. A task often consists of a set/sequence of actions to fulfill the allotted tasks, extremely just one action. And, a task can be as broad and loosely defined as, serving the elder, or as precisely defined as making tea. The scope of the task is defined by its use. As shown later, when discussing hierarchy of tasks and according contexts, this variability of scope of task is an important element of context.

### 4.1. Hierarchy of Tasks and Contexts

In this section, we attempt to generalize this notion to cover a wide range of possible task definitions. Generally speaking, a task can be as broadly defined as, serving the elderly, or as narrowly defined as making tea. Similarly, how different task-specific context information will apply depends on the task itself. The actual scope of the task is therefore defined by its intended use, and thus to generalize, we can further define a set of related tasks and their dependency on context information in a hierarchical manner as shown in Figure 2.

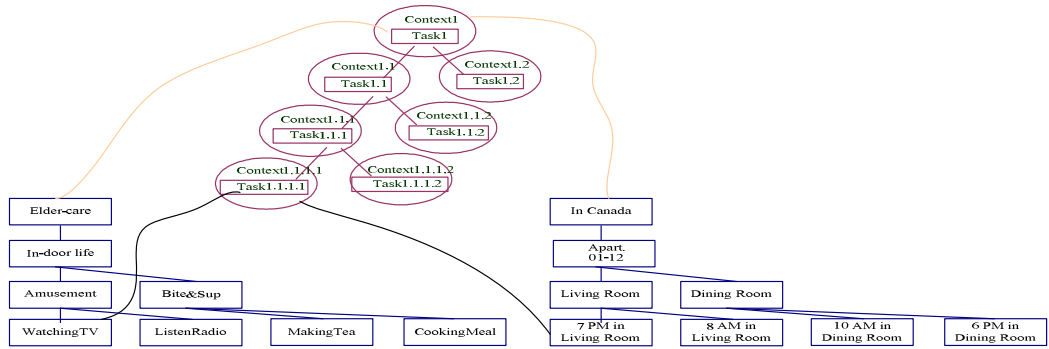

Figure 2. Hierarchy of Contexts and Tasks

To elaborate, in Figure.2, Task1 (e.g. "eldercare"), for instance, can be refined by Task1.1 (e.g. "activities at home") and Task1.1 can also be further refined by Task1.1.1 (e.g. "relaxation"). At the lowest level there are real tasks which cannot be decomposed, such as Task1.1.1.1 (e.g. "watching TV"). As shown in Figure.2, three types of tasks exist in the hierarchy: an overall and generic task (root node), composite tasks (intermediate nodes) and atomic tasks (leaf nodes).

On the other hand, the context relevant to individual tasks can be similarly defined using the task hierarchy. Hence, in Figure.2, Context1 (e.g. "in Canada"), would be relevant to Task1. Similarly, Context 1.1.1.1 (e.g. "7:00 pm in living room") is related to Task1.1.1.1. In summary, whenever a task is decomposed into more objective sub-tasks, the related contexts will similarly be more and more specific with a sub-task automatically inherits the context of its parent tasks.

## 4.2. Formalization of the Context-Driven Task Model

We model tasks and their relations on the top of the context-driven task hierarchy explained above, where each can be further decomposed into a set of sub-tasks (in the case of a composite task), or in the case of atomic task, a set of sequential activities. A task can be described by a union of the following vocabulary:

Task-ID (TI): a unique identifier of a task in a pervasive computing application;

Task-Name (TN): a string to distinguish a task and easy to understand for a user;

Condition (C): a set of preconditions, or context information, that must be met before the task can be performed. The condition is specified in the form of parameters.

Priority (Pr): this field denotes the importance and exigency of a task to further facilitate the execution, suspension and re-scheduling of tasks at runtime. For tasks that have the same priority their relative importance will be determined by the priority of their respective parent-tasks.

Task-Contract (TC): this is a crucial element in our task definition. Task contract has two roles: one is to discovery necessary resources and services for the task; the other is to organize and guide the steps of executing a task. In summary, based on our explanation above, each task is represented with a nested 5-tuple as follows:

$$T= (TI, TN, C, Pr, TC)$$

For example, the Gettingup_Assisting task can be denoted as following:

> ***TaskName***: *Task_Gettingup_Assisting;*
> ***TaskId***: *Task1.1.1.3;*
> ***Condition***: *User_Location; User_state; Time;*
> ***Priority***: *3*
> ***Task-Contract***: *Task_Contract_GettingUp_Assisting;*

Furthermore, the main purpose of task contract (TC) is to organize pervasive resources and services to a composite or atomic task. A task contract can be further defined with the following elements:

Contract-ID (CI): unique identifier of a contract, equal to Task-ID;

Contract-Name (CN): to distinguish a contract that is easy to understand for a user;

Parent-Task (PT): to describe the relationship among different tasks, especially the parent-son tasks;

Requirement (R): to express the necessary material, services (abstract of software or devices). Notice that the Requirement field is very different to Condition field in task model for two reasons. First, Condition depicts a situation surrounding a task (i.e. What), but Requirement describes the resources will be utilized in performing a task (i.e. How). Second, Condition contains an encoding of all relevant aspects about the current environment while Requirement only contains a tailored description of the relevant resources within the environment. To some extent, Condition is more generic than Requirement.

Procedure (P): this field can contain two different sets of values depending on whether the task is composite or atomic nature. In the case of an atomic task, Procedure field will include a sequence of actions that will be executed by some services associated to either some automatic devices and/or software. On the other hand, if this TC belongs to a composite task, then this field will contain information of its "leaf" tasks instead.

Using the above explanations therefore, a task contract can be defined as shown in Fig.5. In general, however, we say a TC is denoted by a 5-tuple, for example, the task contract of Task1.1 in Fig. 3 can be denoted as:

$$TC1.1= (CI1.1, CN1.1, PT1.1, R1.1, P1.1)$$

We will give a specfic example a task contract in section 6 combined the task execution mechanism.

International Journal of UbiComp (IJU), Vol.2, No.1, January 2011

## 4.3. Construction of Context-Driven Task Model Using OWL

We here address how to express the context information using OWL based on the proposed multi-contextual task model. The element of Condition mainly includes two kinds of descriptive information. The first is information on the user (knowledge of habits, emotional state, physiological conditions, etc.), and the other is information related to physical environment, such as location, time, weather, infrastructure (surrounding resources for computation, communication, etc.), and physical conditions (noise, light, pressure, etc.). All the contexts are represented as first-order predicate calculus. Each entity in a pervasive environment can be described in the form of Predicate (subject, value) as following:

subject   $S^*$: set of subject names, e.g., a person, a location or an object.

Predicate   $V^*$: set of predicate names, e.g., is located in, has status, etc.

value   $O^*$: set of all values of subjects in $S^*$, e.g., bedroom, open, close, etc.

For example, Location (Zhang, Livingroom) - Zhang is located in the living room; Temperature (Livingroom, 28) - the temperature of the living room is 28ºC; Status (door3, close) – the No.3 door's (living room door) status is close.

The structures and properties of context predicates are described in an ontology which may include descriptions of classes, properties and their instances. The ontology is written in OWL as a collection of RDF triples with each statement in the form of (subject, verb, object), where subject and object are ontology's objects or individuals, and predicate is a property relation defined by the ontology.

In our model, all the entities in real world are represented as ontology instances and associated properties (entity mark-ups) that can be easily interpreted by applications. Real-world entities often originate from diverse sources, leading to dissimilar approaches to generating different mark-ups. Some of the contexts (e.g., name of a person, gender, and mobile-phone) have relatively slow rates of change. Mark-ups of these contexts are usually generated by users. For example, we provide a JavaScript application that allows users to online create profiles based on the ontology class User. Following example shows the context mark-up that describes nihongbo.

```
<User rdf:ID="nihongbo">

<user_status rdf:datatype="http://www.w3.org/2001/XMLSchema#string"

>normal</user_status>

<studentOf rdf:resource="#zhouxingshe"/>

<family_name rdf:datatype="http://www.w3.org/2001/XMLSchema#string"

>Ni</family_name>

<given_name rdf:datatype="http://www.w3.org/2001/XMLSchema#string"
```

```
            >Hongbo</given_name>

            <gender rdf:datatype="http://www.w3.org/2001/XMLSchema#string"

            >M</gender>

            <mobile_phone rdf:datatype="http://www.w3.org/2001/XMLSchema#string"

            >13891866713</mobile_phone>

        </User>
```

On the other hand, some other contexts (e.g., location, current time, noise level, and door status) are usually provided by hardware or software sources. The marking up of these contexts needs to be performed by automated programs due to the high rate of change, e.g., the RFID indoor location system that tracks users' location by detecting the presence of body-worn tags. When nihongbo enters DiningRoom, the RFID sensor detects his presence and composes the context mark-up as described below:

```
<User rdf:about="#nihongbo"><locatedIn rdf: about ="#DiningRoom"/> </User>
```

Since each OWL instance has a unique URI, entities mark-ups can link to external definitions through these URIs. For example, http://www.dcel.nwpu.edu.cn/SemanticSpace#NiHong-bo and http://www.dcel.nwpu.edu.cn/SemanticSpace#DiningRoom refers to the user and a room respectively.

## 5. CONTEXT-DRIVEN TASK REASONING WITH CBR

In this section we will infer the user's task with the current contexts using CBR approach. In our previous work, we have adopted the rule-based reasoning technique for task inference [19], but as for a smart home, any researchers can't consider all the situations the users will encounter, thus it's impossible for them to design a complete rule-based system. Meanwhile, CBR has its own advantages, as CBR not only reuses previous cases, but also stores new cases for future reference. If no rule in database matches the new case, CBR system will store the new case as a rule. Since the researcher can't envision all situations in smart home in advance, it's difficult to predict an inhabitant's next activity, and thus CBR may be a promising approach for reasoning in smart home environment. To use CBR for reasoning context-driven task, we should consider four important issues: case representation, similarity measurement, case retrieval and solution reuse. Due to the space constraint of the paper, we will focus on case representation and similarity measurement.

### 5.1 Case Representation

Case representation is the first step for CBR. In the field of CBR, case representation is related not only to knowledge storage but also to knowledge reasoning. According to the topic of the paper, we will use CBR to reason the inhabitant's task, and now, the important question to answer is: What are the common aspects between task contexts in task modelling and cases in case-based reasoning as we want to express the of users' intention? First, a context and a case seem to be similar in their definition as a description of aspects of a situation. Of course, contexts obtain a more punctual sense concerning time and cases on the other hand may also represent entire processes. But the use of attribute-value pairs to organize cases is the most common form of representation [20], and is also a powerful "language" for describing both contexts and task models.

Generally speaking, a case includes some key attributes, as seen in the follows:

Case = (Case ID, Sulution description, Problem description)

Meanwhile, in our task models, the dependent contexts in the vector 'Condition' can be described as:

Task = (Task ID, Task Name, User ID, Location, Time, Temprature, Noise,...)

Thus, to represent case using the contexts in our task models, we can transform the parameters in task model into case components.Fig.1 illustrates our approach of representing a case by mapping the problem solution pairs of a case description to context-driven attributes pairs. The problem description complies with the description of the dependent context to a task. As this context description covers the several dimensions as mentioned, the case's problem description is a multi-dimensional vector of complex attributes.

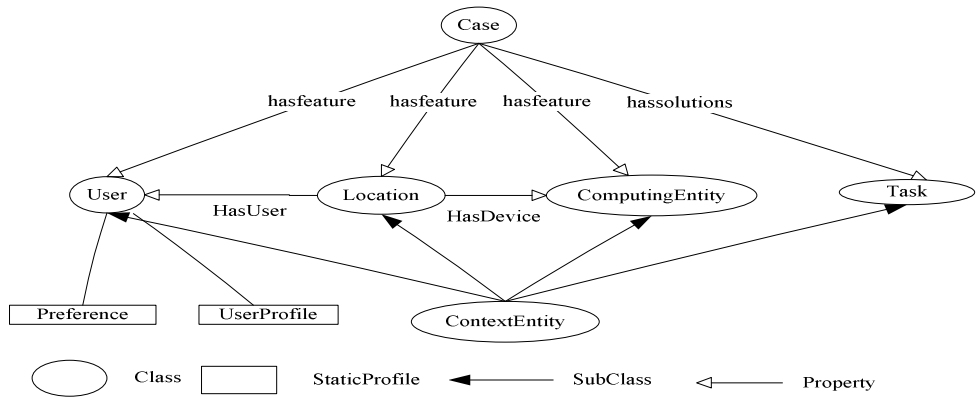

Figure 3. Case Representation with Task Model Using Ontology

As shown in Figure.3, each task has a corresponding case, and such a case can be organized in a table form and stored in a database. In the initial stage, the table consists of some key attributes: such as User ID, Location, Time, etc.. Besides the main attributes, more other attributes can also be added according to the contexts related to a task.

After representing the cases with the contexts, the reasoning procedure is equal to the task inferring with CBR. The structures and properties of context predicates are described in an ontology which may include descriptions of classes, properties and their instances. The ontology is written in OWL as a collection of RDF triples, each statement being in the form (subject prop, object), where subject and object are ontology's objects or individuals, and predicate is a property relation defined by the ontology. Table 1 describes the storage of a complete case and related contexts in the  database, and this case (CaseID is 142) means that the user (NHB) is executing the Bathing task in the BathRoom_30 at 8:40, and it is stored in the case base with the form of RDF triples, as seen in the Figure 4.

Table 1. Case representation with RDF tripels

| Subject | Prop | Obj |
|---|---|---|
| NHB | User_Locatedin | BathRoom_30 |
| NHB | Time | 8:40 |
| BathRoom_30 | HasDevice | Shower |
| Shower | Entity_state | On |
| BathRoom_30 | HasDevice | Lamp-2 |
| Lamp-2 | Entity_state | On |
| BathRoom_30 | Door-State | Off |
| NHB | User_Task | Bathing |

Figure 4. Case Storage with RDF Triples in Database

## 5.2 Similarity Measurement

As mentioned in the above section, a case is organized as a table, and each column is equal to a context attribute, and the whole table equals the condition of a task. So, the similarity is defined on two levels: local similarity and global similarity. The generalized distance can be used to estimate similarity.

The key idea is to match the case parameters to the value dynamically aggregated from the environmental and user contexts. Specifically, let us define each Condition to be a context tuple, i.e. $C=<c_1,c_2,\ldots,c_n>$, $c_1\ldots c_n$ are a set of context attributes, as well the corresponding parameters in one case table. In an actual system, Context tuple values are sampled periodically. In these tuples, there may be many types of attribute values (both numerical value and Boolean). Each kind of attribute values has its own similarity calculation method, which can be expressed in a general form as follows:

$$dis(v(c_i), v'(c_i)) = \frac{|v(c_i) - v'(c_i)|}{dom(c_i)} \quad (1)$$

where $c_i$ means a context attribute, $v(c_i)$ is an expected value, $v'(c_i)$ is the real-time value, and $dom(c_i)$ means the maximal difference of two values $v(c_i)$, $v'(c_i)$. Obviously, for any attribute $c_i$, the value $dis(v(c_i), v'(c_i))$ is within[0,1].

The table (Condition vector) similarity is the combination of all evaluated attribute similarities, and we can revise the typical distance formula to measure the table distance.

Manhattan Distance:
$$dis(T(c), T'(c)) = \sum_{i=1}^{n} \left| \frac{v(c_i) - v'(c_i)}{dom(c_i)} \right| \quad (2)$$

Here, each attribute has the equal contribution for the similarity, in fact, we should further take into consideration that diverse attributes have different contribution to the Condition's overall similarity, and an attribute weight is used for this purpose, for example, location, and time can have a higher weight than others. So the overall similarity can be evaluated as follows:

$$dis(T(c), T'(c)) = \sum_{j} wj \, dis(v(cj), v'(cj)) \quad (3)$$

where $\sum_{j=1} wj = 1$. The range of $dis(T(c), T'(c))$ is [0, 1], a value of zero means perfect match and 1 meaning complete mismatch.

For example, we can define a 'noon break' task as following:

*NoonBreak_Task= (Task_1122, NoonBreak, User_Ni, InBed, Noon, Door_Closed)* As to this task, the corresponding contexts includes: user location, time and the status of the door. Specifically, we can detect the user location with pressure sensor on the mattress, and the time will be provided by the system clock, and the status of the door is checked by the angle sensor on the top of the door. Assume that the user sleeps in the bed at 12:15 with the door left unlocked; we can calculate several contexts with diverse method as following Table 2.

Table 2. Context Similarity Calculating in the Noonbreak-task

| Property | Detecting Method | Default Weight | Similarity |
|---|---|---|---|
| Location | number of activated/total (30) | 0.466 | 2030=0.667 |
| Time | real time in [12:00-13:30] | 0.277 | 1 (time = 12:15) |
| Door_Status | detected angle/180 | 0.257 | 120/180 |

The overall similarity can be calculated with the equation (3), i.e. 0.466*0.667+0.277*1+0.257*0.67=0.76, so we can infer the probability of the Noonbreak task is 76%, that is, the user is likely sleeping in the bed for noon-break.

International Journal of UbiComp (IJU), Vol.2, No.1, January 2011

## 5. CBR Framework and Performance Evaluation

To integrate CBR in CDTOM, we have designed a CBR-based reasoning sub-system (as seen in Figure 5), which is to discover a context-driven task in a smart home. In CBR sub-system, the original context data, such as time, location, and temperature is generated by sensors, and then sent to the case representation module. After structured, the real-time context data is filtered into the case table, and the similarity can be calculated between the real-time case and template casefrom the task-base. The best match case is retrieved from task-base database, that is, the result is the inhabitant's task.

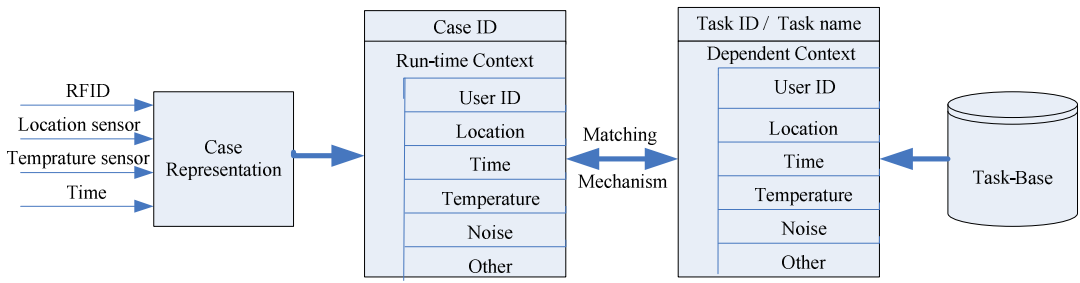

Figure 5. CBR Sub-system Framework

According to the application scenarios in smart home, we have conducted the experiments on a windows XP platforms, the system configuration is: CPU: P4-2.4G, RAM: 512Mb, IDE: Eclipse3.0.1 +JDK1.4.2. We applied the Protégé + MySql to compile and store the context data, and the experiment results are as follows:

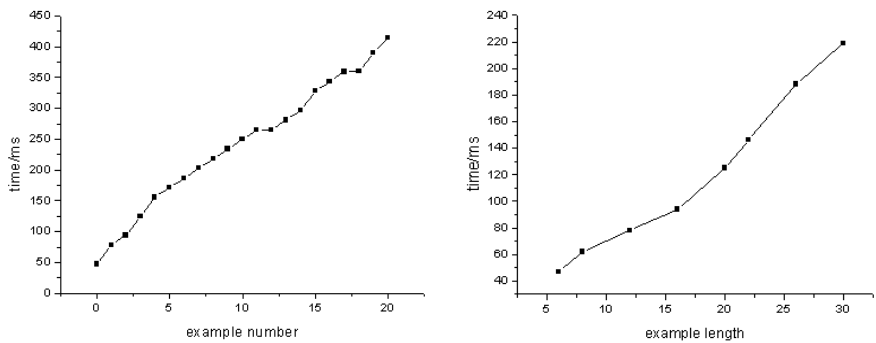

Figure 6. The Experiments result of CBR Reasoning

As can be seen in Figure 6 (a), the size of case is the key factor for the system performance, and the minimum time is 78ms (for one case), the maximum time is 415ms (for 20 cases). Since we have classified the cases into 6 types according to the user place, there will not be too many cases in a specific location, so we think the performance of the reasoning system is acceptable in smart home environment. Furthermore, as can be seen in Figure. 6 (b), the dimension of attributes is another important factor, and the matching duration is proportional to the size of case and the number of attributes.

## 6. TASK EXECUTION AND SERVICE MAPPING

Having defined the task contract in the previous section, we shall now focus on the steps involving in executing a task through the procedure field defined as part of the task contract. As mentioned, for an atomic task, the procedure field within the task contract will contain one or more discrete action steps. These steps specify the sequential actions to be taken during task execution, along with a number of discrete events that may happen during the execution (see Figure 7). Conceptually, a procedure can be regarded as a finite automaton, with each state corresponds to a step during task execution, and an event which might occur during the execution. An event can come from outside user or from inside devices, sensor and software. Briefly, Fig. 4 shows a graphical representation of the relationships among the various context-driven task model components elaborated so far. The first table is the set of tasks, and the second one is a current task inferred from the context information. The third table is a corresponding task contract, and the actual procedure (i.e. actions) of performing the task illustrated in the last table. Finally, on the extreme right is the set of events that the current action in the procedure is either producing or waiting for.

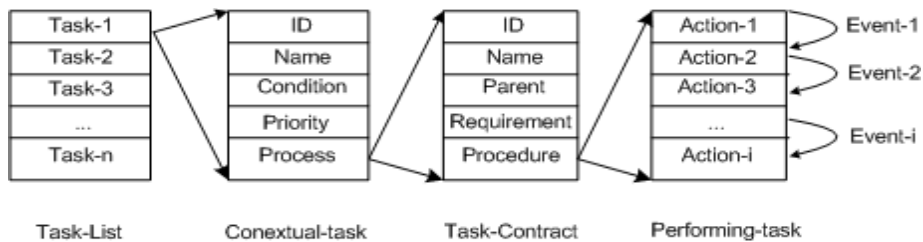

Figure 7. Relationships between Task Execution and Task Contract

As known, all actions in a task will be performed by diverse services invoked by the events and located in different devices, such as smart phone, smart TV, etc. In CDTOM, we adopt the Event-Condition-Action (ECA) mechanism to implement the mapping between actions and services.

ECA rules were originally proposed in the database community in order to provide traditional database systems with the capability of event-driven, instantaneous response [21, 22]. Since the ECA approach offers a means of the concise and distinct descriptions of reactive behaviors, it has been adopted in a variety of application fields, In recent years, some authors have introduced ECA into pervasive computing for context-aware applications [23][24][25].in this paper, we will use ECA rule to invoke the relative services according to the current contexts and bridge the gap between abstract services and the specific ones. In the proposed framework of CDTOM, a service device is invoked by various event sources and service logical providers that, respectively, offer event triggering and specific services. That is, service devices are assumed to interact each other via the events generated via publish/subscribe mechanism and the specific service invocations that use request/response interaction model based on Active MQ.

Following the structure of traditional ECA rules, each rule can be described as 3-tuple, i.e. $R = <E, C, A>$, that is, each rule consists of events, conditions, and actions.

The event is a triggering notification message that can be one of four primitive event types, namely (i) *internal event* if it is generated by a system variables (global/local), (ii) *context event* if it is delivered from a context source, (iii) *time event* that occurs after certain time has elapsed, and finally (iv) *service event* that is generated at the point of invoking a service of a device.

The condition is a boolean expression that must be satisfied for some action of a device to occur. It is defined by use of event variables contained in an event message.

Finally, the action represents an instruction performing a specific service by a device, for example, display a video, or turn down the air-conditioner.

In CDTOM, the grammar of ECA rule language can be described as following:

Rules = "rules" [Targets] "{"{Rule} "}".

Targets = "for" Entity Or Template {"," Entity Or Template}.

Rule = "When" Event ["If" Condition] "THEN DO"Action

Where, Rules denote the ECA rule set for a given task, that is, a task can be executed by one or more rules, and the rules can be executed in three modes: sequence, choice and loop. Targets represent the object entities, usually including users, actuators and service providers; Rule denotes the specific item in the rule set.

In order to trigger a specific service, we encapsulate each ECA rule into a message, and also utilize the publish/subscription mechanism for message communicating as seen in Figure 8.

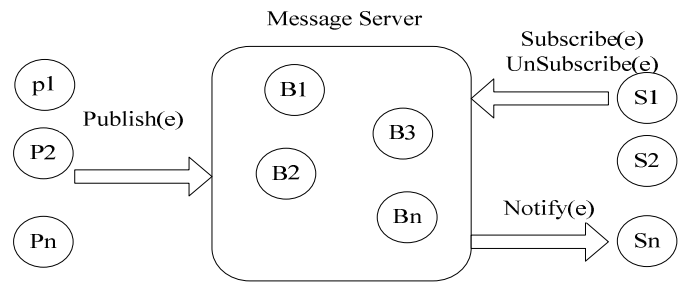

Figure 8. The Pub/Sub Mechanism in Message Server

Where P denotes the context provider, S stands for the Service actuator, E represents the Event.

Now, we discuss a simple example to explain the procedure of ECA rule mechanism. In a smart home, when the temperature in the bedroom is over 30℃ or the humidity is over 80, the inhabitant will feel uncomfortable, and then the air-conditioner should be started to adjust the atmosphere in bedroom. Figure.9 denotes the format of message from a temperature sensor, and the subscription from the air-conditioner service with an ECA rule, separately.

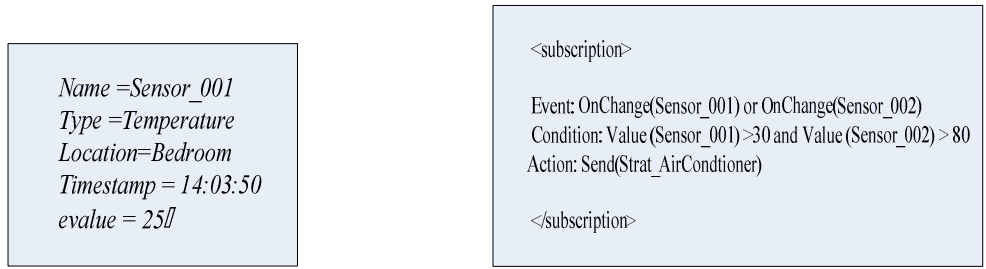

Figure 9. Description of Message and ECA rule in CDTOM

As an example, the task of Gettingup_Assisting can be exetued by the ECA-rule embedded task contract as following:

```
Contract ID: 1.1.1.1
Contract Name: Task_Contract_Gettingup_Assisting;
Parent Task:
  HealthCare;
Requirement:
  Light, Blind, Air_Conditioner,PDA;
Procedure:
  Begin {
      rules for <Sensor_Provider> {
          When Bed_ pressure_Sensor.triggered() {Then DO
          <GettingUp_Message_Dispatcher>.GettingUpMessage:Send();
          }
          } //Rule for Gettingup_Aware service
          rules for <PDA> {
          When User_A_isGettingUP() {Then DO
          <Weather_Forecast_Displayer>.Weather_Forecast:Play();
          <Schedule_Reminder>.Remind:Query();
          <Schedule_Reminder>.Remind:Display();
          }
          } //rule for Weather_Forecast & Schedule_Reminder services
          rules for <Air_Conditioner> {
          When User_A_isGettingUP()
          {if Temperature_bedroom.CurrentValue<20 &&
          humidity_bedroom.CurrentValue>60
          Then DO
          <Air_Conditioner_Controller>.Air_Conditioner:Start()
          }
          }     } End
```

Figure 10. The ECA rule Embedded Task Contract

In summary, The ECA mechanism provides a high level structure for CDTOM that proactively react upon context changes. It has been devised in order to decouple context concerns from reaction (communication and service usage) concerns, under control of the context-driven task model. In this way, the ECA scheme enables the coordination, configuration and cooperation of pervasive services within CDTOM.

## 7. SYSTEM IMPLEMENTATION

CDTOM is one of the key projects under the National High Technology Research and Development Program of China that aims to provide the middleware infrastructure for pervasive computing environment. We are making continuous research progress on the development and improvement of various components in CDTOM. These components are progressively integrated into a system prototype for performance evaluation and functional demonstration of the middleware infrastructure. In this section, we will report the current prototype system named Task Supporting System (TSS) in our lab.

Figure 11 shows the system architecture of the TSS, we used 3 desktop PCs for the system servers to manage the contexts/tasks, messages and services, in context/task server, Each PC has an Intel Core2 2.6GHz CPU and 2GB main memory running Windows XP. We implemented the context management and task processor using the MySQL and Jena2 Semantic Web Toolkit, and message pub/sub mechanism using ActiveMQ. Moreover, we developed the service discovery and management using OSGi 3.0 and UPnP SDK v1.01.

Besides, we deployed the various sensors in our workplace to detect environment contexts, including the temperature, humidity, light intensity, etc. the user's location is an important context, we used RFID and UWB tags respectively to monitor the coarse and refined position of the user. Furthermore, the user can interact with the TSS using smart phone or PDA and get the services from the customized smart appliances, such as Digital TV, Air Condition, etc.

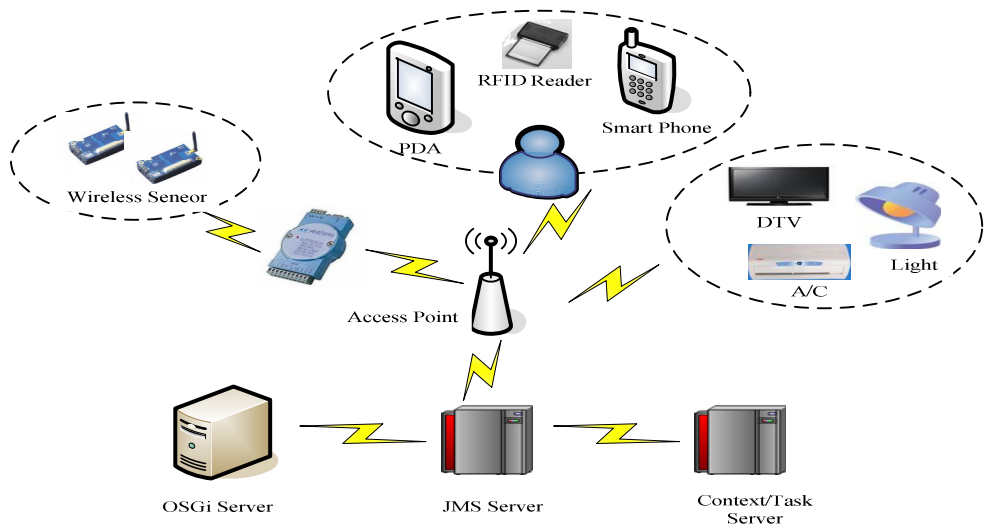

Figure 11. The Architecture of TSS Using CDTOM

Till now, the TSS can provides adaptive and personalized task assisting applications, such as environment adjustment, medicine taking reminder, scheduler management, telephone forwarding, as seen in the following screen-shots in Figure.12.

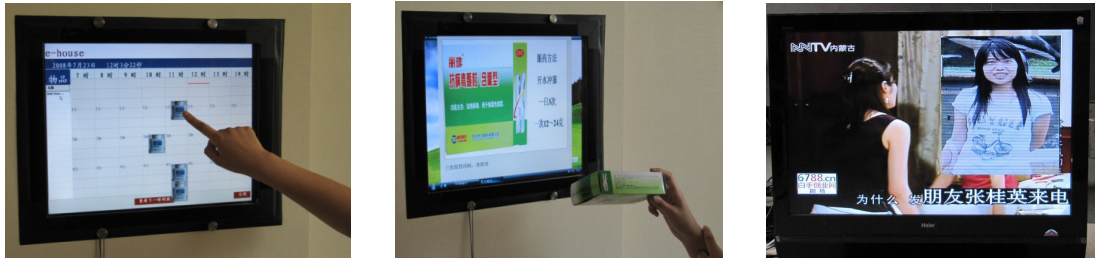

Figure 12. Screenshots of Task assisting Applications

## 8. CONCLUSIONS

We present in this paper our design and implementation of a context-driven task-oriented middleware for pervasive homecare environment. CDTOM is used to support various pervasive tasks for the elders in smart home. The main strength of CDTOM is its separation of tasks and services, which would allow for greater flexibility of changing the tasks without changing the services and vice-verse, and also hiding the complexity of compositing various services in pervasive environment from the users. It can deliver task assisting applications to end-users in an automatic and adaptive manner, and enable the elder to live safely and conveniently.

Our key contributions in this paper are: (i) the concept and model of Context-driven task is proposed by defining the relationship between users' tasks and ambient contexts, (ii) the Case-Based Reasoning technique is deployed and the descriptive model of the case, the algorithm of the weight calculating, the algorithm of case similarity calculating are thoroughly presented, (iii)the ECA mechanism is introduced for mapping the action sequences and corresponding services, (iv) based on the application background of elder-care in the smart home, the paper implements a platform called TSS using CDTOM, which integrates task management, information processing, and context management to support users' task in pervasive environments.

The current prototype system is still under developing and we will invite some elders to evaluate the system and assess the usability issue. There are also some outstanding technical issues to be solved. We are working on a universal model to expand smart home to an open smart space, semi-or unsupervised method of inferring user's task, and also some improvements in service mapping methods.

## ACKNOWLEDGEMENTS

This work is being supported by the Fonds Nature et Technologies, MELS Program, Quebec, Canada, and partially supported by the National High Technology Research and Development Program of China under Grant No. 2009AA011903.

## AUTHORS

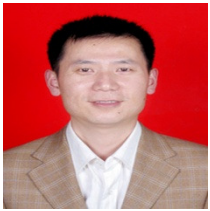

Hongbo Ni received his Ph.D. in Computer Science from Northwestern Polytechnical University, China. His is currently a postdoctoral research fellow in the Department of Computer Science, Sherbrooke University, Canada. His research interests includes pervasive computing,embedded systems and assistive technology.

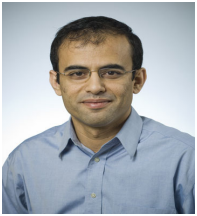

BESSAM ABDULRAZAK finished his Doctor of Computer Sciences from the TELECOM SudParis (ex. INT: Institut National des Télécommunications) and University of Evry, France. His is currently an Assistant Professorin the Department of Computer Science sherbrooke university, Canada. His research interests are in the areas of Ambient Intelligence, Assistive Robotic, and Human Machine Interaction.

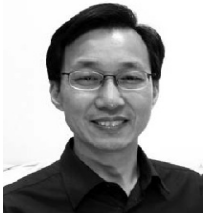

Daqing ZHANG is a professor in Ambient Intelligence and Pervasive System Design at Institute TELECOM SudParis, France. He received his Ph.D. from University of Rome "La Sapienza" and University of L'Aquila, Italy. Dr. Daqing's research interests include pervasive/ubiquitous computing, service-oriented computing, and context-aware systems.

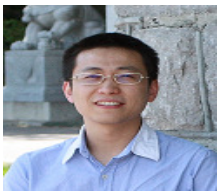

Shu Wu received the MSc degree in computer science in 2007. He is currently working toward the PhD degree in the Department of Computer Science, University of Sherbrooke, Quebec, Canada. His research interests include the area of data mining and pervasive computing.